\documentclass{article}

\usepackage{arxiv}

\usepackage[utf8]{inputenc} 
\usepackage[T1]{fontenc}    
\usepackage{hyperref}       
\usepackage{url}            
\usepackage{booktabs}       
\usepackage{amsfonts}       
\usepackage{nicefrac}       
\usepackage{microtype}      
\usepackage{lipsum}		
\usepackage{graphicx}
\usepackage{natbib}
\usepackage{doi}

\usepackage{amsmath}

\title{
The Gradient of Health Data Privacy
}

\author{
Baihan Lin, PhD\\
	Berkman Klein Center For Internet \& Society, Harvard Law School, Cambridge, MA 02138 \\
    Departments of AI, Psychiatry, and Neuroscience, Icahn School of Medicine at Mount Sinai, New York, NY 10029 \\
	\texttt{blin@law.harvard.edu} \\
}

\renewcommand{\headeright}{Baihan Lin}

\hypersetup{
pdftitle={The Gradient of Health Data Privacy},
pdfsubject={q-bio.NC, q-bio.QM},
pdfauthor={Baihan Lin},
pdfkeywords={First keyword, Second keyword, More},
}

\begin{document}
\maketitle

\begin{abstract}
In the era of digital health and artificial intelligence, the management of patient data privacy has become increasingly complex, with significant implications for global health equity and patient trust. This paper introduces a novel ``privacy gradient'' approach to health data governance, offering a more nuanced and adaptive framework than traditional binary privacy models. Our multidimensional concept considers factors such as data sensitivity, stakeholder relationships, purpose of use, and temporal aspects, allowing for context-sensitive privacy protections. Through policy analyses, ethical considerations, and case studies spanning adolescent health, integrated care, and genomic research, we demonstrate how this approach can address critical privacy challenges in diverse healthcare settings worldwide. The privacy gradient model has the potential to enhance patient engagement, improve care coordination, and accelerate medical research while safeguarding individual privacy rights. We provide policy recommendations for implementing this approach, considering its impact on healthcare systems, research infrastructures, and global health initiatives. This work aims to inform policymakers, healthcare leaders, and digital health innovators, contributing to a more equitable, trustworthy, and effective global health data ecosystem in the digital age.
\end{abstract}




\section{Introduction}

In the age of artificial intelligence and big data, health information has become an invaluable resource for medical research, personalized healthcare, and public health policy. However, current approaches to health data privacy, often based on binary models of either complete privacy or full accessibility, fail to capture the nuanced nature of health information and its varied uses. This paper proposes a novel ``privacy gradient'' approach to health data management, offering a more flexible and context-sensitive framework for protecting patient privacy while maximizing data utility.

The intersection of law, technology, and healthcare has created complex challenges for policymakers \citep{terry2009s}. As legal scholars engage their battles with the implications of rapidly evolving health technologies, computer scientists are developing increasingly sophisticated systems for data management and analysis. Yet, current health data privacy paradigms often leave both groups unsatisfied: legal experts find existing frameworks too rigid to address complex real-world scenarios, while technologists struggle to implement systems that can adapt to the nuanced requirements of healthcare privacy.

This paper introduces the concept of a health data privacy gradient, drawing inspiration from architectural principles such as the ``intimacy gradient'' \citep{alexander2018pattern} and legal theories of contextual integrity \citep{nissenbaum2004privacy}. We argue that by conceptualizing privacy as a spectrum rather than a binary state, we can develop more adaptive legal frameworks and technological solutions that better align with the complex realities of modern healthcare.

\begin{figure}[tb]
\centering
    \includegraphics[width=.8\linewidth]{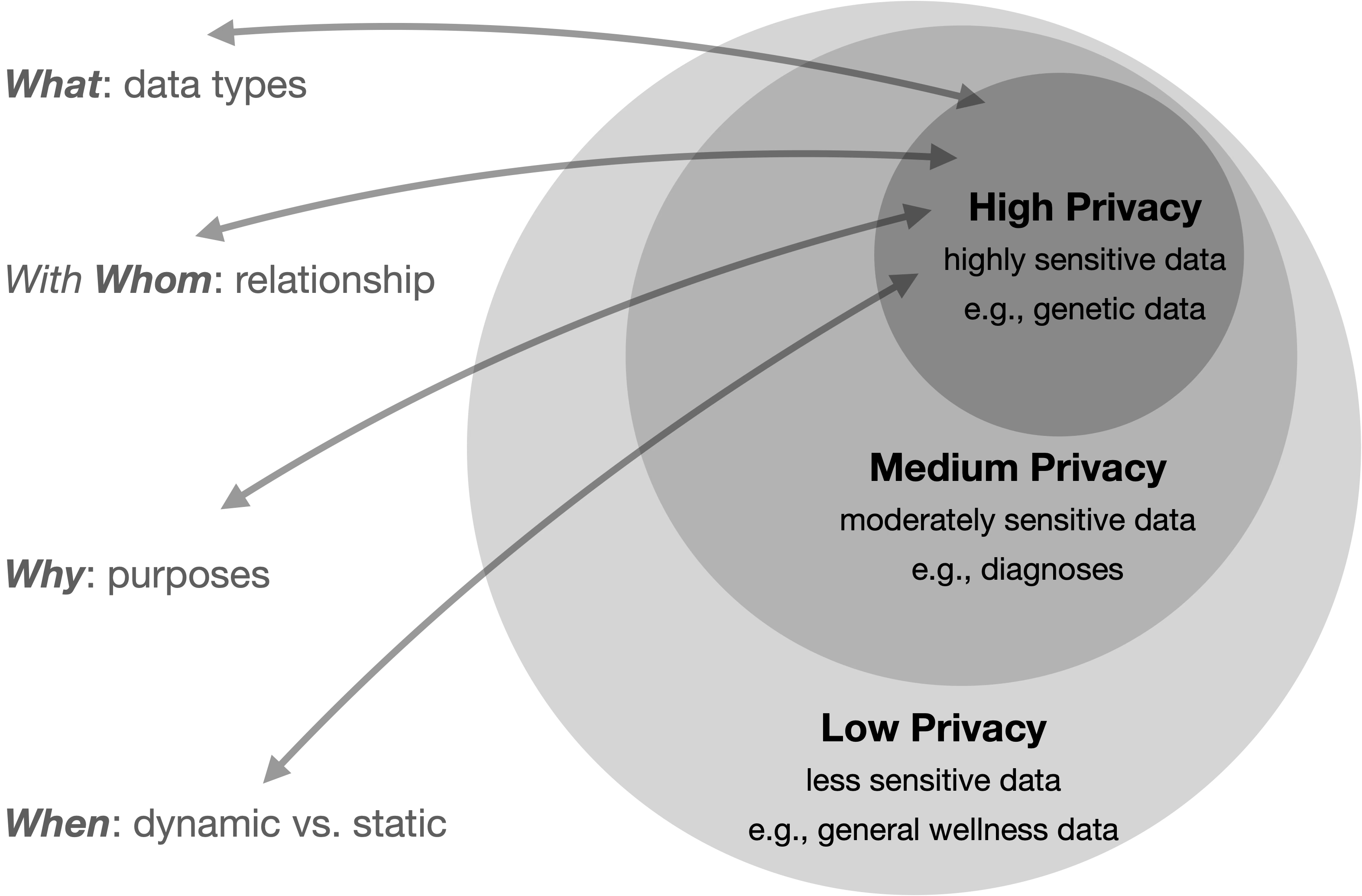}
\caption{\textbf{Conceptual representation of the health data privacy gradient.} Arrows pointing inward and outward represent the dynamic nature of data sensitivity based on context.
}\label{fig:concept}
\end{figure}

Our approach considers multiple factors in determining appropriate levels of privacy protection, including data sensitivity, the relationship between the data subject and user, the purpose of data use, and temporal aspects (Figure \ref{fig:concept} and Table \ref{tab:pg_dimension}). Through case studies spanning adolescent health, integrated care, clinical trials, and genomic research, we demonstrate how this approach can address complex privacy challenges that are difficult to resolve with traditional models.

This paper aims to stimulate dialogue between policymakers, legal scholars, and technologists, encouraging interdisciplinary collaboration to develop more nuanced and effective approaches to health data privacy in the AI era. We conclude by discussing the potential impacts of this approach and outlining key areas for future policy development and research.

\begin{table}[tb]
      \caption{ \textbf{Examples of Privacy Levels Across Different Dimensions}  
      }
      \label{tab:pg_dimension} 
      \centering
 \begin{tabular}{ | l | p{0.25\linewidth} | p{0.25\linewidth} | p{0.25\linewidth} |}
 \hline
\textbf{Dimension} & \textbf{Low Privacy} & \textbf{Medium Privacy} & \textbf{High Privacy} \\ \hline
{Data Sensitivity} & General wellness (e.g., step count) & Diagnoses or chronic condition data & Genetic predisposition data \\ \hline
{Relationship} & Public health official & Treating physician & Patient themselves \\ \hline
{Purpose} & Population-level statistics & Personalized treatment planning & Exploratory genetic research \\ \hline
{Temporal Aspect} & 5-year-old aggregated data & Recent individual health record & Real-time biometric data \\ \hline
\end{tabular}
\end{table}

\section{Background and Related Work}

The concept of privacy in healthcare has a rich history, evolving from ancient principles of medical confidentiality to today's complex legal and technological frameworks. This section will explore this evolution and examine current approaches to health data privacy, setting the stage for our proposed gradient model.

\textbf{Historical Perspective on Privacy in Healthcare.}
The notion of privacy in healthcare dates back to ancient times, with the Hippocratic Oath serving as one of the earliest codifications of medical confidentiality. The oath states, ``\textit{What I may see or hear in the course of the treatment or even outside of the treatment in regard to the life of men, which on no account one must spread abroad, I will keep to myself, holding such things shameful to be spoken about}'' \citep{oath1995hippocratic}. This principle of confidentiality has remained a cornerstone of medical ethics for over two millennia.

As healthcare systems modernized and became more complex, the need for more formal privacy protections became apparent. In the United States, this led to the development of the Health Insurance Portability and Accountability Act (HIPAA) in 1996 \citep{act1996health}. HIPAA established national standards for the protection of individuals' medical records and other personal health information, marking a significant shift towards a legal framework for health data privacy.

\textbf{Current Privacy Models in Digital Health.}
The digital transformation of healthcare has introduced new challenges and opportunities for data privacy. Current approaches to managing health data privacy typically fall into two main categories:

\textit{Role-Based Access Control (RBAC)}: This model restricts system access to authorized users based on their roles within an organization \citep{sandhu1998role}. In healthcare settings, RBAC is often used to ensure that medical professionals only have access to the patient data necessary for their specific roles.

\textit{Consent-Based Models}: These approaches focus on obtaining explicit consent from patients for the use of their data. The General Data Protection Regulation (GDPR) in the European Union, for example, places a strong emphasis on consent and gives individuals significant control over their personal data \citep{gdpr2016general}.

While these models have their merits, they often struggle to capture the nuanced and context-dependent nature of health data privacy. As Solove points out, privacy is not a singular concept but a plurality of related issues \citep{solove2005taxonomy}. This complexity is particularly evident in healthcare, where the sensitivity of information can vary dramatically based on context.

\begin{figure}[tb]
\centering
    \includegraphics[width=.8\linewidth]{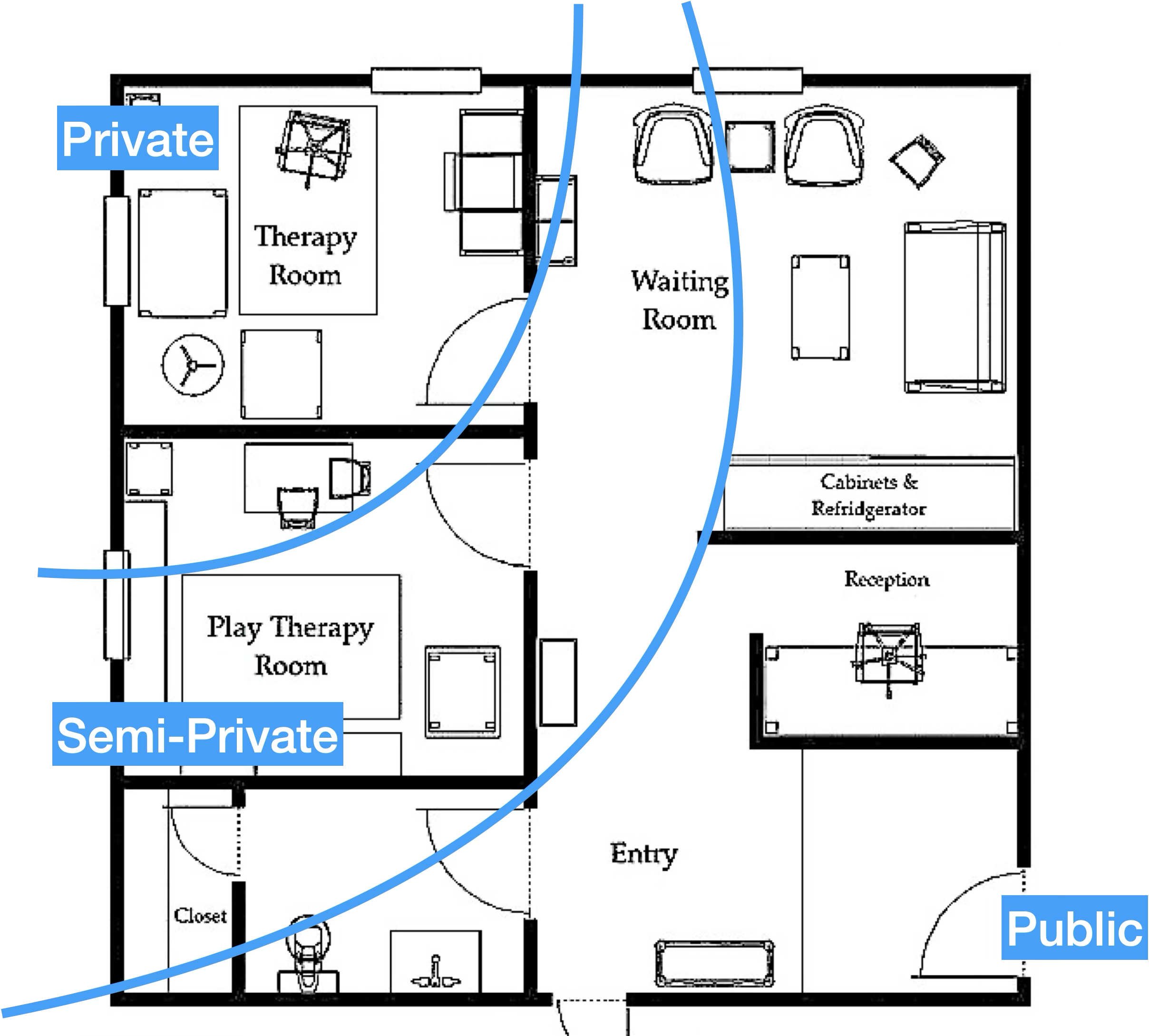}
\caption{\textbf{Architectural representation of the intimacy gradient in a healthcare setting.} Shown is a floor plan with an intimacy gradient from public spaces (e.g., reception area) to semi-private areas (e.g., examination rooms) to private spaces (e.g., counseling rooms). 
}\label{fig:floorplan}
\end{figure}

\textbf{The Architectural Concept of ``Intimacy Gradient''.}
To address these limitations, we draw inspiration from fields outside of healthcare. The concept of an ``intimacy gradient'' was introduced by architect Christopher Alexander in his seminal work ``A Pattern Language'' \citep{alexander2018pattern}. Alexander proposed that buildings and towns should be designed with a spectrum of spaces, ranging from very public to very private. This gradient allows for a natural flow between different levels of intimacy and privacy (Figure \ref{fig:floorplan}).

While originally conceived for physical spaces, the intimacy gradient concept has profound implications for digital privacy. Just as a well-designed building provides appropriate spaces for different levels of privacy, a well-designed health information system should offer a spectrum of privacy options tailored to the sensitivity of the data and the context of its use.

\textbf{Contextual Integrity Theory.}
Building on this idea of context-dependent privacy, Helen Nissenbaum's theory of contextual integrity provides a valuable framework for understanding privacy in the digital age \citep{nissenbaum2004privacy}. Nissenbaum argues that privacy is not about secrecy or control over information, but about the appropriate flow of information based on contextual norms.

In the healthcare domain, contextual integrity suggests that the appropriateness of data sharing depends not just on the type of data, but on the roles of the sender and recipient, the context in which the information is disclosed, and the terms under which the disclosure occurs. This theory aligns closely with our proposed gradient approach, emphasizing the need for flexible, context-aware privacy models.






\textbf{Synthesis and Gap Analysis.}
While each of these approaches and technologies offers valuable insights into health data privacy, they often operate in isolation.
Our proposed gradient of health data privacy seeks to synthesize these diverse approaches into a cohesive framework. By combining the flexibility of the intimacy gradient, the context-awareness of contextual integrity theory, and the technological capabilities of modern privacy-preserving computation, we aim to create a more holistic and adaptable approach to health data privacy.

This gradient approach addresses a critical gap in current privacy models: the need for a framework that can dynamically adjust privacy protections based on the full context of data use, including the type of data, the relationships between parties, the purpose of data use, and temporal factors. In the following sections, we will elaborate on this gradient concept and explore its potential implementations and implications for both legal frameworks and technological systems in healthcare.

\section{The Health Data Privacy Gradient}

The privacy gradient concept represents a paradigm shift in how we approach health data privacy. Instead of viewing privacy as a binary state—where data is either private or public—we propose a continuous spectrum of privacy levels that can be dynamically adjusted based on context. This approach allows for more nuanced and flexible management of health data, better aligning with the complex realities of modern healthcare and research.

\textbf{Defining the Privacy Gradient.}
The privacy gradient can be conceptualized as a multidimensional continuum along which health data can be positioned and repositioned based on various factors. This gradient ranges from highly restrictive privacy settings to more open access, with numerous intermediate states.

As illustrated in Figure \ref{fig:concept}, the privacy gradient is not a simple linear scale but a multidimensional space where different factors interact to determine the appropriate level of privacy protection for a given piece of health data in a specific context. Overall, we can consider it to reside in a 3d surface with axes representing data sensitivity, relationship proximity, and purpose specificity. A specific set of axes values defines a curved surface representing the privacy level, with different colors indicating varying degrees of privacy protection. 

\textbf{Key Dimensions of the Gradient.}
Now we have established the conceptual framework, the privacy gradient is practically shaped by several key dimensions, each of which contributes to determining the appropriate level of privacy protection (exemplified in Table \ref{tab:pg_dimension}):

\begin{enumerate}
    \item \textit{Data Sensitivity}: This dimension considers the inherent sensitivity of the health data. For example, genetic data or mental health records might be considered highly sensitive, while general wellness information might be less sensitive.
    \item \textit{Relationship to Data Subject}: This dimension takes into account the relationship between the data subject (the patient) and the potential data user. A treating physician might have a closer relationship and thus potentially greater access than a researcher or public health official.
    \item \textit{Purpose of Data Use}: The intended use of the data plays a crucial role in determining its position on the privacy gradient. Data used for direct patient care might be more accessible than data used for secondary research purposes.
    \item \textit{Temporal Aspects}: The time factor can significantly impact privacy considerations. Recent or real-time health data might require stronger protections than historical data.
\end{enumerate}

\textbf{Illustrative Scenarios Across the Gradient.}
To better understand how the privacy gradient functions in practice, let's consider three scenarios that illustrate different points along the gradient:

\textit{{Scenario 1. High Privacy: Genetic Predisposition Data}.}
Alice undergoes genetic testing which reveals a predisposition to a rare genetic disorder. This information is highly sensitive and is placed at the high privacy end of the gradient.

\begin{itemize}
    \item Data Sensitivity: Very High (genetic data)
    \item Relationship: Limited to Alice and her genetic counselor
    \item Purpose: Strictly for Alice's personal health management
    \item Temporal Aspect: Lifelong relevance
\end{itemize}
  
In this scenario, the data would be subject to the strictest privacy protections, with access limited to Alice and her designated healthcare providers. Any use of this data for research would require explicit consent and strong anonymization techniques.

\textit{{Scenario 2. Medium Privacy: Chronic Condition Management}.}
Bob has diabetes and uses a continuous glucose monitor that syncs data to his smartphone. This data occupies a middle ground on the privacy gradient.

\begin{itemize}
    \item Data Sensitivity: Moderate (chronic condition data)
   \item Relationship: Bob, his endocrinologist, and his primary care physician
   \item Purpose: Ongoing condition management and treatment adjustment
   \item Temporal Aspect: Recent and ongoing data collection
   \end{itemize}

Here, the privacy settings would allow for sharing with Bob's care team and could potentially be used (with consent) for population-level diabetes research in an anonymized form.

\textit{{Scenario 3. Low Privacy: General Wellness Information}.}
Carol uses a fitness tracker to monitor her daily step count and heart rate. This general wellness data sits at the lower end of the privacy gradient.

\begin{itemize}
    \item Data Sensitivity: Low (non-medical wellness data)
    \item Relationship: Carol and potentially her fitness coach or primary care provider
    \item Purpose: Personal fitness tracking, general health monitoring
    \item Temporal Aspect: Ongoing collection, but individual data points less critical
\end{itemize}
   
This data might be more freely shared, for instance, with Carol's fitness applications or for anonymized public health research on activity levels.

These scenarios demonstrate how the privacy gradient can adapt to different types of health data and usage contexts, providing appropriate levels of protection without unnecessarily restricting beneficial data use.

\textbf{Dynamic Nature of the Privacy Gradient.}
It's crucial to emphasize that a data point's position on the privacy gradient is not static. As contexts change, so too can the privacy level. For instance, if Carol's fitness tracker data showed sudden, concerning changes in her heart rate, its sensitivity might increase, moving it higher on the privacy gradient and potentially triggering alerts to her healthcare provider.

This dynamic aspect of the privacy gradient aligns with the concept of contextual integrity proposed by Nissenbaum \citep{nissenbaum2004privacy}. It recognizes that appropriate information flow is contextual and that privacy norms can shift based on changing circumstances.

\textbf{Challenges and Considerations.}
While the privacy gradient offers a more nuanced approach to health data privacy, it also presents challenges:

A multidimensional, dynamic privacy model is inherently more \textit{complex} than binary privacy settings. This could potentially lead to confusion for users and implementation difficulties for system designers. Achieving a \textit{standardized} understanding of privacy levels across different healthcare systems and jurisdictions could be challenging. Balancing the need for dynamic, context-based privacy adjustments with \textit{user control} and transparency is a significant consideration. Lastly, ensuring that a gradient approach aligns with existing \textit{legal compliance} frameworks like HIPAA and GDPR will require careful consideration and potentially legislative updates.

Despite these challenges, we believe that the privacy gradient approach offers significant benefits in terms of flexibility, contextual appropriateness, and the potential to unlock valuable data use while maintaining robust privacy protections.

In the next section, we will explore how this conceptual model could be implemented technically, considering both current technologies and potential future developments.

\section{Technical Implementation of a Privacy Gradient Model}

\begin{table}[tb]
      \caption{ \textbf{Examples of Access Control Factors Across the Privacy Gradient}  
      }
      \label{tab:access} 
      \centering
 \begin{tabular}{ | l | l | l | l | }
 \hline
\textbf{Access Control Factor} & \textbf{Low Privacy} & \textbf{Medium Privacy} & \textbf{High Privacy} \\ \hline
{User Role} & Public Health Researcher & Treating Physician & Patient \\ \hline
{Data Type} & Aggregated Statistics & Individual Health Record & Genetic Data \\ \hline
{Access Purpose} & Population Health Study & Direct Patient Care & Personal Review \\ \hline
{Access Location} & Any & Hospital Network & Secure Terminal Only \\ \hline
{Time Constraint} & Business Hours & 24/7 & Scheduled Appointments Only \\ \hline
\end{tabular}
\end{table}

Translating the conceptual model of a privacy gradient into a functioning technical system presents both challenges and opportunities. This section will explore potential approaches to implementing a privacy gradient in health informatics systems, discussing key technologies and methodologies.

\textbf{Data Classification and Tagging.}
The foundation of a privacy gradient system is a robust method for classifying and tagging health data. This process must be both granular enough to capture the nuances of different data types and flexible enough to adapt to changing contexts.


   Machine learning algorithms can be employed to automatically classify incoming health data based on its content, source, and context. For example, natural language processing (NLP) techniques could be used to analyze clinical notes and assign initial privacy classifications \citep{meystre2008extracting}.
   Utilizing semantic web technologies, health data can be tagged with rich metadata that describes not just the data type, but also its context, potential uses, and privacy implications. The Fast Healthcare Interoperability Resources (FHIR) standard provides a foundation for such semantic tagging in healthcare IT systems \citep{bender2013hl7}.

   As the context of data use changes, the system must be capable of dynamically reclassifying data. This could involve periodic re-evaluation of data classification based on new information or triggers from system events.

\textbf{Dynamic Access Control Mechanisms.}
Traditional role-based access control (RBAC) systems are too rigid to fully implement a privacy gradient. Instead, we propose a dynamic access control system that takes into account multiple factors to determine data access in real-time.

\textit{Attribute-Based Access Control} (ABAC) extends beyond RBAC by considering a wide range of attributes about the data, the user, and the context when making access decisions \citep{hu2017attribute}. This aligns well with our multidimensional privacy gradient concept.

Building on ABAC, a more relevant approach in our case would be \textit{Context-Aware Access Control} (CAAC), a context-aware system would consider factors such as time, location, device type, and current system state when making access decisions. For example, access to certain data might be granted only during office hours or from secure locations.

Incorporating the principle of purpose specification from privacy laws, access control decisions would consider the declared purpose for data access. We call this approach \textit{Purpose-Based Access Control} (PBAC). This could be implemented through a system of purpose declarations that are matched against allowed purposes associated with each data element. The access control factors can be summarized in Table \ref{tab:access}.

\textbf{Privacy-Preserving Techniques.}
To enable useful data processing while maintaining privacy, especially for data at the higher end of the privacy gradient, advanced privacy-preserving computation techniques can be employed:

   Differential privacy adds calibrated noise to dataset queries, allowing for meaningful statistical analysis while protecting individual privacy \citep{dwork2006differential}. This technique could be particularly useful for allowing research access to sensitive health data.
   Homomorphic encryption allows computations to be performed on encrypted data without decrypting it \citep{gentry2009fully}. This could enable secure processing of highly sensitive health data in untrusted environments, such as cloud computing platforms.
Secure Multi-Party Computation (MPC)  allows multiple parties to jointly compute a function over their inputs while keeping those inputs private \citep{yao1982protocols}. In healthcare, this could facilitate collaborative research on sensitive data across institutions without sharing the raw data.

\textbf{User Interfaces for Gradient-Based Privacy Management.}
Effective implementation of a privacy gradient system requires user interfaces that can convey complex privacy settings in an intuitive manner:

\begin{enumerate}
    \item \textit{Visual Privacy Dashboards}:
   Interactive dashboards could use color gradients and other visual cues to represent the current privacy state of different data types. Users could adjust privacy levels using slider controls or similar intuitive interfaces.
   \item  \textit{Contextual Privacy Notifications}:
   The system should provide just-in-time notifications to users about privacy implications of their actions. For example, when a physician attempts to access sensitive patient data, a notification could explain the reason for the elevated privacy level and request additional confirmation.
   \item \textit{Privacy Setting Templates}
   To simplify management of complex privacy settings, the system could offer pre-configured templates for common scenarios (e.g., ``Research Study Participant", ``Chronic Disease Management"). Users could then customize these templates as needed.
\end{enumerate}


\textbf{Interoperability and Standards.}
For a privacy gradient approach to be widely adopted, it must be compatible with existing health IT standards and support interoperability across systems.

   The \textit{Fast Healthcare Interoperability Resources} (FHIR) standard could be extended to include privacy gradient metadata, allowing for seamless exchange of privacy-tagged health data between compliant systems.
   \textit{Blockchain} technology could be used to create immutable audit trails of privacy setting changes and data access events, enhancing transparency and accountability \citep{azaria2016medrec}.
   Building on the \textit{OpenID Connect} standard, a privacy gradient-aware authentication system could communicate user attributes and purpose declarations to enable context-aware access control decisions.

\textbf{Challenges on the Technology Side.}
While these technologies offer promising avenues for implementing a privacy gradient, several challenges remain:

The computational cost of real-time, context-aware access decisions and privacy-preserving computation techniques could impact system performance with \textit{performance overhead}.
Balancing the complexity of gradient-based privacy with the \textit{usability} need for intuitive user interfaces is an ongoing challenge.
As health data volumes grow, maintaining fine-grained privacy controls at \textit{scale} will require innovative approaches to data management and processing.
Even with advanced techniques, the risk of \textit{privacy leakage} through inference attacks or combination of multiple data sources remains a concern.


In the next section, we will explore the legal and ethical implications of implementing a privacy gradient approach in health informatics.

\section{Legal and Ethical Implications}

The implementation of a privacy gradient approach in health informatics raises significant legal and ethical considerations. This section explores how existing legal frameworks might adapt to this new paradigm and examines the ethical implications of a more nuanced approach to health data privacy.

\subsection{Adapting Existing Legal Frameworks}

\textbf{{Reinterpreting HIPAA for Gradient Privacy}.}
The Health Insurance Portability and Accountability Act (HIPAA) in the United States is a cornerstone of health data privacy regulation. However, HIPAA's binary approach to data (either Protected Health Information or not) doesn't align perfectly with a gradient model.

\begin{table}[tb]
      \caption{ \textbf{Potential Adaptations of HIPAA Principles to a Privacy Gradient Model}  
      }
      \label{tab:hipaa} 
      \centering
 \begin{tabular}{ | l | p{0.25\linewidth} | p{0.5\linewidth} |}
 \hline
\textbf{HIPAA Principle} & \textbf{Current Interpretation} & \textbf{Gradient Privacy Adaptation}  \\ \hline
{Minimum Necessary} & Access limited to minimum necessary information & Dynamic determination of ``minimum necessary'' based on gradient position \\ \hline
{Authorization} & Binary consent for data use & Granular, purpose-specific authorizations aligned with gradient levels \\ \hline
{De-identification} & Safe Harbor or Expert Determination methods & Contextual de-identification aligned with gradient position \\ \hline
{Breach Notification} & Based on risk of compromise to PHI & Risk assessment considering gradient position of affected data \\ \hline
\end{tabular}
\end{table}

Adapting HIPAA to a gradient model would require reinterpreting key principles (Table \ref{tab:hipaa}):

 The ``\textit{minimum necessary}'' standard could be dynamically determined based on the data's position on the privacy gradient.
\textit{Authorization} for data use could become more granular, allowing patients to consent to specific uses aligned with different gradient levels.
\textit{De-identification standards} might need to be reimagined as a spectrum rather than a binary state, with the level of de-identification required varying based on the data's gradient position.

Legal scholar Mark Rothstein has argued for a more nuanced approach to health data privacy that considers context and sensitivity, which aligns well with our gradient model \citep{rothstein2010deidentification}.

\textbf{{GDPR and the Right to Privacy}.}
The European Union's General Data Protection Regulation (GDPR) offers a more flexible framework that could potentially accommodate a gradient approach. The GDPR's principles of data minimization, purpose limitation, and storage limitation align well with the dynamic nature of a privacy gradient.

However, implementing a gradient approach under GDPR would require careful consideration of several aspects:

\textit{The right to erasure} (``right to be forgotten'') might need to be reinterpreted in a gradient context, where data might move to higher privacy levels rather than being completely erased.
The concept of ``\textit{legitimate interest}'' as a basis for data processing could be aligned with different levels of the privacy gradient.
\textit{Data Protection Impact Assessments} (DPIAs) could incorporate gradient-based risk assessments.

\subsection{Ethical Considerations}

\textbf{{Patient Autonomy and Informed Consent}.}
A privacy gradient approach has the potential to enhance patient autonomy by offering more granular control over health data. However, it also raises questions about the nature of informed consent in a complex, dynamic privacy environment.


Ethicist Ruth Faden and Tom Beauchamp has argued that true informed consent requires not just disclosure of information, but also understanding and voluntariness \citep{faden1986history}. In a gradient privacy system, ensuring that patients fully understand the implications of their privacy choices becomes even more critical and challenging.

\textbf{{Balancing Individual Privacy with Public Health Needs}.}

The privacy gradient approach offers new possibilities for balancing individual privacy rights with broader public health interests. For instance, during a public health emergency, certain types of health data might temporarily shift to a lower privacy level to facilitate rapid response and research.

However, this flexibility also raises ethical concerns. As public health ethicist James Childress and colleagues notes, there is often tension between public health measures and other moral considerations such as individual liberty and privacy \citep{childress2002public}. A gradient approach would need to carefully navigate this tension.

\textbf{{Health Equity and Non-Discrimination}.}
The implementation of a privacy gradient system must consider issues of health equity and potential discrimination. There's a risk that complex privacy systems could disadvantage certain populations, such as those with lower health literacy or limited access to technology.

Moreover, the ability to more finely tune access to health data could potentially be misused for discriminatory purposes. Safeguards would need to be in place to prevent the privacy gradient from being used to unfairly target or exclude certain individuals or groups.

\textbf{Challenges in Standardization and Interoperability.}
There are additional non-technical challenges to establish the standardization and interoperability of privacy gradient.
First, health data often needs to flow across \textit{jurisdictional boundaries}, whether for multi-national research projects or for providing care to traveling patients. A privacy gradient approach would need to be standardized across different legal jurisdictions to ensure interoperability while respecting local privacy laws.

The technical implementation of a privacy gradient would need to \textit{align} closely with legal and ethical standards. This requires close collaboration between technologists, legal experts, and ethicists to develop standards that are both technically feasible and legally compliant.

Implementing a privacy gradient approach would require robust \textit{governance} structures to ensure \textit{accountability}. This might involve the creation of new oversight bodies or the expansion of existing ones, such as Institutional Review Boards (IRBs) in the academic insitutions, to handle the complexities of gradient-based privacy decisions.

\subsection{Potential Legal and Ethical Benefits}

Despite these challenges, a privacy gradient approach offers several potential benefits from a legal and ethical perspective:
First, it offers \textit{enhanced proportionality} to existing legal and ethical frameworks. By allowing privacy protections to be tailored to the specific context and sensitivity of the data, a gradient approach could better align with legal principles of proportionality.

Second, it improves \textit{transparency}. A well-implemented gradient system could provide greater transparency about how health data is used and protected, potentially increasing trust in health information systems.

Third, by allowing more nuanced control over data access, a gradient approach could facilitate valuable health research while still maintaining strong protections for sensitive data.

Last but not least, it \textit{empowers patients}. Giving patients more granular control over their health data aligns with ethical principles of respect for persons and could enhance patient engagement in their own healthcare.

As we move towards implementing privacy gradient systems, ongoing dialogue between technologists, legal scholars, ethicists, healthcare providers, and patients will be crucial to navigate these complex issues and develop systems that are both technically robust and ethically sound.
In the next section, we will examine several case studies that illustrate how a privacy gradient approach might be applied in real-world healthcare scenarios.

\section{Case Studies}

To better understand the practical implications and potential benefits of a privacy gradient approach, we will examine four diverse scenarios in healthcare. These case studies will demonstrate how a gradient approach to privacy can address complex challenges that are difficult to resolve with traditional binary privacy models.

\textbf{Parental Access to Adolescent Health Records.}
A 16-year-old patient, Sarah, is seeking treatment for depression. She wants to keep certain aspects of her mental health information private from her parents, but her parents argue they need full access to her records to make informed decisions about her care.

\textit{Traditional Approach:}
In many jurisdictions, parents have the right to access their minor children's complete medical records, with some exceptions for sensitive information like reproductive health. This can lead to adolescents avoiding necessary care due to privacy concerns.

\begin{table}[tb]
      \caption{ \textbf{Gradient-Based Access to John's Health Data Across Care Team}  
      }
      \label{tab:john_access} 
      \centering
 \begin{tabular}{| l | l | l | l |}
 \hline
\textbf{Data Type} & \textbf{ Primary Care} & \textbf{Endocrinologist} & \textbf{Psychiatrist} \\ \hline
{Diabetes Diagnosis} & Full Access & Full Access & Limited Access \\ \hline
{Depression Diagnosis} & Full Access & Limited Access & Full Access \\ \hline
{Medication List} & Full Access & Full Access & Full Access \\ \hline
{Therapy Notes} & No Access & No Access & Full Access \\ \hline
{Integrated Care Plan} & Full Access & Full Access & Full Access \\ \hline
\end{tabular}
\end{table}

\textit{Privacy Gradient Approach:}
The privacy gradient approach would determine different levels of parental access to an adolescent's health record, from full access (general health information) to no access (confidential mental health notes):

\begin{itemize}
    \item General Health Information (Low Privacy): Parents have full access to general health information, vaccination records, and physical exam results.
    \item Mental Health Diagnosis (Medium Privacy): Parents are informed of the diagnosis but don't have access to detailed therapy notes.
    \item Therapy Session Notes (High Privacy): These remain confidential between Sarah and her therapist, with exceptions for imminent safety risks.
\end{itemize}

The system could dynamically adjust access based on Sarah's age, evolving capacity, and specific health needs. As Sarah approaches adulthood, the gradient could shift to give her more control over her data.

\textit{Legal and Ethical Considerations:}
This approach aligns with the concept of the ``mature minor doctrine'' recognized in some jurisdictions \citep{coleman2013legal,sigman1991exploration,cherry2010parental,coleman2021adolescent}. It balances the parents' need to make informed decisions with the adolescent's growing autonomy and right to privacy, potentially encouraging more open communication between adolescents and healthcare providers.

\textbf{Mental Health Data in Integrated Care Settings.}
John is receiving treatment for both diabetes and depression. His care team includes his primary care physician, an endocrinologist, and a psychiatrist. The challenge is to share relevant information among the team while respecting the sensitive nature of mental health data.

\textit{Traditional Approach:}
Mental health information often receives special protection under privacy laws, which can lead to siloed care and missed opportunities for holistic treatment.

\begin{figure*}[tb]
\centering
    \includegraphics[width=\linewidth]{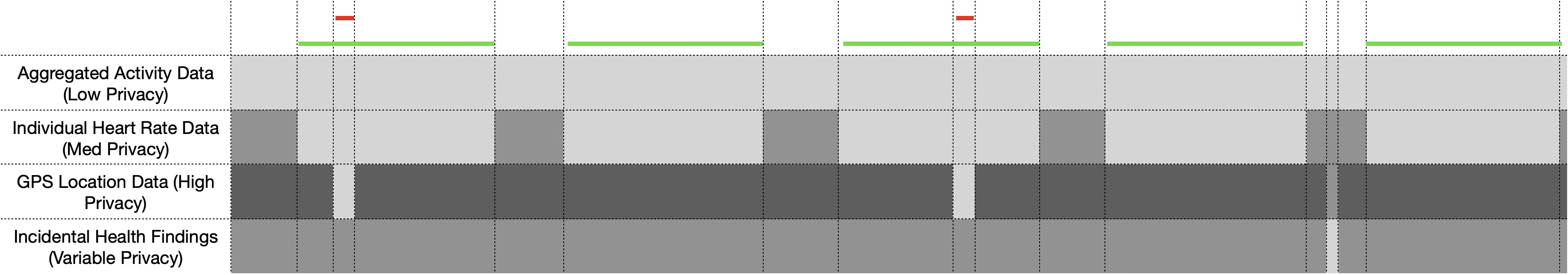}
\caption{\textbf{Dynamic Privacy Gradient in a Clinical Trial with Wearable Devices.} Shown here is a timeline showing different phases of the clinical trial with corresponding privacy levels for different types of data. Green bars marked normal working hours, and red bars for medical emergencies. The privacy levels are represented by a color gradient (the lighter the color, the less sensitive the data is), with certain data types becoming more or less private at different stages of the trial.
}\label{fig:pg_timeline}
\end{figure*}

\textit{Privacy Gradient Approach:}
The system could allow John to adjust these settings, for instance, temporarily elevating access during a health crisis. An example gradient-based access to John's health data across care team is shown in Table \ref{tab:john_access}, with some notable access factor decisions include:

\begin{itemize}
    \item Diagnoses (Medium Privacy): All team members have access to both diagnoses to enable integrated care.
    \item Medication List (Low Privacy): Shared across all providers to prevent drug interactions.
    \item Detailed Mental Health Notes (High Privacy): Accessible only to the psychiatrist, with a summary available to other providers if John consents.
    \item Integrated Care Plan (Medium Privacy): Accessible to all team members, focusing on how the conditions interact without disclosing sensitive details.

\end{itemize}

\textit{Legal and Ethical Considerations:}
This approach aligns with the principles of integrated care while respecting the heightened privacy concerns around mental health data. It could potentially improve care coordination without compromising patient confidentiality, addressing the challenges highlighted by researchers in mental health integration \citep{bauer2019effectiveness}.

\textbf{Wearable Device Data in Clinical Trials.}
A pharmaceutical company is conducting a clinical trial for a new heart medication. Participants are asked to wear fitness trackers to monitor their heart rate and activity levels continuously.

\textit{Traditional Approach:}
Participants typically sign a broad consent form at the beginning of the trial, granting researchers access to all collected data.

\textit{Privacy Gradient Approach:}
As in Figure \ref{fig:pg_timeline}, the privacy gradient approach adopts a dynamic privacy assignment over different phases of the clinical trial, each with changing privacy levels for different types of data:

\begin{itemize}
    \item Aggregated Activity Data (Low Privacy): Continuously shared with researchers throughout the trial.
    \item Individual Heart Rate Data (Medium Privacy): Shared in real-time during active trial periods, but restricted during off-hours.
    \item GPS Location Data (High Privacy): Collected but only accessed in the event of a medical emergency.
    \item Incidental Health Findings (Variable Privacy): If the device detects a potential health issue unrelated to the study, the participant is notified first and can decide whether to share this information with the research team or their personal physician.
\end{itemize}

The system could allow participants to temporarily elevate privacy levels (e.g., during personal events) and provide clear audit trails of data access.

\textit{Legal and Ethical Considerations:}
This approach addresses concerns raised by ethicists about the continuous monitoring involved in some clinical trials \citep{nebeker2017ethical}. It provides more granular control to participants, potentially increasing willingness to participate in trials while ensuring data integrity for researchers.

\textbf{Genomic Data Sharing for Research.}
A large-scale genomics research project aims to identify genetic factors contributing to rare diseases. Participants are asked to share their genetic data, which has implications not just for them but also for their biological relatives.

\textit{Traditional Approach:}
Genetic data is typically treated as highly sensitive, with stringent access controls. This can limit the potential for valuable research.

\begin{table}[tb]
      \caption{ \textbf{Privacy Gradient Levels for Different Types of Genomic Data}  
      }
      \label{tab:genome_access} 
      \centering
 \begin{tabular}{ | l | l | l |}
 \hline
\textbf{Data Type} & \textbf{Privacy Level} & \textbf{Access Conditions}  \\ \hline
{Presence/Absence of Specific Variants} & Low & Widely accessible to researchers \\ \hline
{Anonymized Genomic Sequences} & Medium & Accessible to approved research projects \\ \hline
{Full Identifiable Genome} & High & Restricted access, requires specific consent \\ \hline
 {Familial Linkage Information} & Variable & Depends on consent of family members \\ \hline
\end{tabular}
\end{table}

As summarized in Table \ref{tab:genome_access}, the privacy gradient levels can be determined by different types of genomic data under different access conditions:

\begin{itemize}
    \item De-identified Genetic Variants (Low Privacy): Shared broadly with researchers studying specific conditions.
    \item Anonymized Full Genomic Sequences (Medium Privacy): Available to approved research projects, with access logged and audited.
    \item Identifiable Genomic Data (High Privacy): Tightly controlled, requiring explicit consent for each use.
    \item Familial Linkage Data (Variable Privacy): Managed through a dynamic consent process involving multiple family members.
\end{itemize}

The system could allow participants to adjust privacy levels for different parts of their genomic data and receive notifications about how their data is being used.

\textit{Legal and Ethical Considerations:}
This approach addresses some of the complex ethical issues in genomic data sharing identified by scholars \citep{o2021toward}. It balances the potential for groundbreaking research with individuals' right to control their genetic information, while also considering the familial nature of genetic data.

\textbf{Synthesis.}
These case studies demonstrate how a privacy gradient approach can provide more nuanced solutions to complex health data privacy challenges. By moving beyond binary notions of privacy, this approach can:

\begin{enumerate}
    \item Enhance patient autonomy and engagement
\item Facilitate more effective care coordination and research
\item Provide flexibility to adapt to changing circumstances and individual preferences
\item Balance competing interests (e.g., parental rights vs. adolescent privacy, research needs vs. individual control)
\end{enumerate}

However, implementing such a system would require careful consideration of technical feasibility, user understanding, and alignment with legal and ethical frameworks. It would also necessitate ongoing dialogue between stakeholders to refine and adapt the model as new challenges emerge.

In our final section, we will discuss the potential impact of the privacy gradient approach and outline future directions for research and implementation.

\section{Policy Implications and Recommendations}

The privacy gradient approach to health data management represents a paradigm shift with significant policy implications. This section outlines key policy recommendations and explores the potential impacts of implementing this approach.

\textbf{Modernizing Legal Frameworks.}
Current healthcare privacy laws, such as HIPAA in the United States and GDPR in Europe, are based on relatively binary notions of privacy. To accommodate a gradient approach, we recommend:

\begin{enumerate}
    \item Amending HIPAA to incorporate gradient-based privacy levels, allowing for more nuanced control over Protected Health Information (PHI).
    \item Expanding GDPR's data minimization and purpose limitation principles to explicitly support gradient-based access controls.
    \item Developing new legislative frameworks that recognize the multi-dimensional nature of health data privacy, considering factors such as data sensitivity, relationship to the data subject, purpose of use, and temporal aspects.
\end{enumerate}

\textbf{Enhancing Patient Empowerment and Trust.}
To leverage the privacy gradient approach for improved patient engagement:

\begin{enumerate}
    \item Mandate the development of user-friendly interfaces that allow patients to visualize and control their privacy settings easily.
    \item Require healthcare providers to offer privacy education programs, helping patients understand the implications of their privacy choices.
    \item Establish guidelines for transparent reporting of how patient data is used, particularly in research and AI development contexts.
\end{enumerate}

\textbf{Facilitating Research and Innovation.}
To balance privacy protection with the need for data access in research:

\begin{enumerate}
    \item Develop policies that allow for more granular consent processes, enabling patients to share specific types of data for research while maintaining higher privacy levels for other data.
    \item Create regulatory sandboxes to test gradient-based privacy approaches in research settings, allowing for controlled evaluation of their effectiveness.
    \item Establish guidelines for anonymization and de-identification that align with the privacy gradient concept, potentially allowing for more data to be safely used in research.
\end{enumerate}

\textbf{Addressing Implementation Challenges.}
To overcome barriers to adoption:

\begin{enumerate}
    \item Allocate government funding for the development of standardized APIs and protocols for gradient-based privacy systems.
    \item Offer tax incentives or grants to healthcare organizations implementing privacy gradient systems, offsetting the initial costs of adoption.
    \item Mandate interoperability standards that incorporate privacy gradient concepts, ensuring consistent application across different healthcare systems and jurisdictions.
\end{enumerate}

\textbf{International Cooperation.}
Given the global nature of health data and AI development:

\begin{enumerate}
    \item Establish international working groups to develop global standards for gradient-based health data privacy.
    \item Create frameworks for cross-border health data sharing that incorporate privacy gradient principles.
    \item Develop model legislation that countries can adapt to their specific contexts while maintaining international compatibility.
\end{enumerate}

\section{Conclusion and Future Directions}

The privacy gradient approach offers a promising path forward in balancing the need for data access in the AI era with robust privacy protections. By moving beyond binary notions of privacy, it has the potential to enhance patient trust, improve data utility for research and care, and provide more nuanced solutions to complex privacy challenges in healthcare.

Implementing this approach will require concerted effort from policymakers, healthcare providers, technologists, and patient advocates. Key areas for future policy development include:

\begin{enumerate}
    \item Developing comprehensive guidelines for implementing gradient-based privacy systems in healthcare organizations.
    \item Creating certification processes for privacy gradient-compliant systems and organizations.
    \item Establishing ongoing monitoring and evaluation mechanisms to assess the impact of gradient-based privacy approaches on patient trust, data availability for research, and healthcare outcomes.
    \item Exploring the application of privacy gradient concepts in other sectors dealing with sensitive personal data.
\end{enumerate}

As we continue to refine and implement the privacy gradient approach, we have the opportunity to reshape health data governance for the digital age. By embracing this more nuanced and flexible approach to privacy, we can work towards a future where health data drives innovation and improves care while respecting individual privacy rights. The challenge now lies in translating this conceptual framework into concrete policies and practices that can be adopted across the healthcare ecosystem.

\bibliographystyle{apalike}
\bibliography{main}  

\begin{thebibliography}{}

\bibitem[Act, 1996]{act1996health}
Act, A. (1996).
\newblock Health insurance portability and accountability act of 1996.
\newblock {\em Public law}, 104:191.

\bibitem[Alexander, 2018]{alexander2018pattern}
Alexander, C. (2018).
\newblock {\em A pattern language: towns, buildings, construction}.
\newblock Oxford university press.

\bibitem[Azaria et~al., 2016]{azaria2016medrec}
Azaria, A., Ekblaw, A., Vieira, T., and Lippman, A. (2016).
\newblock Medrec: Using blockchain for medical data access and permission management.
\newblock In {\em 2016 2nd international conference on open and big data (OBD)}, pages 25--30. IEEE.

\bibitem[Bauer et~al., 2019]{bauer2019effectiveness}
Bauer, M.~S., Miller, C.~J., Kim, B., Lew, R., Stolzmann, K., Sullivan, J., Riendeau, R., Pitcock, J., Williamson, A., Connolly, S., et~al. (2019).
\newblock Effectiveness of implementing a collaborative chronic care model for clinician teams on patient outcomes and health status in mental health: a randomized clinical trial.
\newblock {\em JAMA network open}, 2(3):e190230--e190230.

\bibitem[Bender and Sartipi, 2013]{bender2013hl7}
Bender, D. and Sartipi, K. (2013).
\newblock Hl7 fhir: An agile and restful approach to healthcare information exchange.
\newblock In {\em Proceedings of the 26th IEEE international symposium on computer-based medical systems}, pages 326--331. IEEE.

\bibitem[Cherry, 2010]{cherry2010parental}
Cherry, M.~J. (2010).
\newblock Parental authority and pediatric bioethical decision making.
\newblock {\em Journal of Medicine and Philosophy}, 35(5):553--572.

\bibitem[Childress et~al., 2002]{childress2002public}
Childress, J.~F., Faden, R.~R., Gaare, R.~D., Gostin, L.~O., Kahn, J., Bonnie, R.~J., Kass, N.~E., Mastroianni, A.~C., Moreno, J.~D., and Nieburg, P. (2002).
\newblock Public health ethics: mapping the terrain.
\newblock {\em The Journal of Law, Medicine \& Ethics}, 30(2):170--178.

\bibitem[Coleman and Rosoff, 2013]{coleman2013legal}
Coleman, D.~L. and Rosoff, P.~M. (2013).
\newblock The legal authority of mature minors to consent to general medical treatment.
\newblock {\em Pediatrics}, 131(4):786--793.

\bibitem[Coleman and Rosoff, 2021]{coleman2021adolescent}
Coleman, D.~L. and Rosoff, P.~M. (2021).
\newblock Adolescent medical decisionmaking rights: Reconciling medicine and law.
\newblock {\em American journal of law \& medicine}, 47(4):386--426.

\bibitem[Dwork, 2006]{dwork2006differential}
Dwork, C. (2006).
\newblock Differential privacy.
\newblock In {\em International colloquium on automata, languages, and programming}, pages 1--12. Springer.

\bibitem[Faden and Beauchamp, 1986]{faden1986history}
Faden, R.~R. and Beauchamp, T.~L. (1986).
\newblock {\em A history and theory of informed consent}.
\newblock Oxford University Press.

\bibitem[GDPR, 2016]{gdpr2016general}
GDPR, G. D. P.~R. (2016).
\newblock General data protection regulation.
\newblock {\em Regulation (EU) 2016/679 of the European Parliament and of the Council of 27 April 2016 on the protection of natural persons with regard to the processing of personal data and on the free movement of such data, and repealing Directive 95/46/EC}.

\bibitem[Gentry, 2009]{gentry2009fully}
Gentry, C. (2009).
\newblock Fully homomorphic encryption using ideal lattices.
\newblock In {\em Proceedings of the forty-first annual ACM symposium on Theory of computing}, pages 169--178.

\bibitem[Hu et~al., 2017]{hu2017attribute}
Hu, V.~C., Ferraiolo, D.~F., Chandramouli, R., and Kuhn, D.~R. (2017).
\newblock {\em Attribute-Based Access Control}.
\newblock Artech House.

\bibitem[Meystre et~al., 2008]{meystre2008extracting}
Meystre, S.~M., Savova, G.~K., Kipper-Schuler, K.~C., and Hurdle, J.~F. (2008).
\newblock Extracting information from textual documents in the electronic health record: a review of recent research.
\newblock {\em Yearbook of medical informatics}, 17(01):128--144.

\bibitem[Nebeker et~al., 2017]{nebeker2017ethical}
Nebeker, C., Harlow, J., Espinoza~Giacinto, R., Orozco-Linares, R., Bloss, C.~S., and Weibel, N. (2017).
\newblock Ethical and regulatory challenges of research using pervasive sensing and other emerging technologies: Irb perspectives.
\newblock {\em AJOB empirical bioethics}, 8(4):266--276.

\bibitem[Nissenbaum, 2004]{nissenbaum2004privacy}
Nissenbaum, H. (2004).
\newblock Privacy as contextual integrity.
\newblock {\em Wash. L. Rev.}, 79:119.

\bibitem[Oath, 1995]{oath1995hippocratic}
Oath, H. (1995).
\newblock The hippocratic oath.
\newblock {\em Am J Med Genet}, 58:187--94.

\bibitem[O’Doherty et~al., 2021]{o2021toward}
O’Doherty, K.~C., Shabani, M., Dove, E.~S., Bentzen, H.~B., Borry, P., Burgess, M.~M., Chalmers, D., De~Vries, J., Eckstein, L., Fullerton, S.~M., et~al. (2021).
\newblock Toward better governance of human genomic data.
\newblock {\em Nature genetics}, 53(1):2--8.

\bibitem[Rothstein, 2010]{rothstein2010deidentification}
Rothstein, M.~A. (2010).
\newblock Is deidentification sufficient to protect health privacy in research?
\newblock {\em The American Journal of Bioethics}, 10(9):3--11.

\bibitem[Sandhu, 1998]{sandhu1998role}
Sandhu, R.~S. (1998).
\newblock Role-based access control.
\newblock In {\em Advances in computers}, volume~46, pages 237--286. Elsevier.

\bibitem[Sigman and O'Connor, 1991]{sigman1991exploration}
Sigman, G.~S. and O'Connor, C. (1991).
\newblock Exploration for physicians of the mature minor doctrine.
\newblock {\em The Journal of pediatrics}, 119(4):520--525.

\bibitem[Solove, 2005]{solove2005taxonomy}
Solove, D.~J. (2005).
\newblock A taxonomy of privacy.
\newblock {\em U. Pa. l. Rev.}, 154:477.

\bibitem[Terry, 2009]{terry2009s}
Terry, N.~P. (2009).
\newblock What's wrong with health privacy?
\newblock {\em J. Health \& Biomedical L.}, 5:1.

\bibitem[Yao, 1982]{yao1982protocols}
Yao, A.~C. (1982).
\newblock Protocols for secure computations.
\newblock In {\em 23rd annual symposium on foundations of computer science (sfcs 1982)}, pages 160--164. IEEE.

\end{thebibliography}

\end{document}